\documentclass[preprint,notoc,letterpaper]{JHEP3}
\usepackage{epsfig}

\def\to{\rightarrow}

\def\te{\tilde e}

\def\tu{\tilde u}

\def\tb{\tilde b}

\def\td{\tilde d}

\def\tst{\tilde t}
\def\ttau{\tilde \tau}
\def\tmu{\tilde \mu}
\def\tg{\tilde g}
\def\tnu{\tilde\nu}

\def\tw{\widetilde W}
\def\tz{\widetilde Z}

\newcommand{\gtsim}{\stackrel{>}{\sim}}

\title{Yukawa coupling unification and non-universal gaugino mediation of
supersymmetry breaking}

\author{Csaba Bal\'azs \\
Department of Physics, Florida State University\\
Tallahassee, FL 32306, U.S.A.\\
E-mail: \email{balazs@hep.fsu.edu}}
\author{Radovan Derm\' \i \v sek \\
Davis Institute for High Energy Physics,
University of California,  \\
Davis, CA 95616, U.S.A. \\
E-mail: \email{dermisek@physics.ucdavis.edu}}

\preprint{FSU-HEP-030319 \\ UCD-03-19}

\abstract{
The requirement of Yukawa coupling unification highly constrains the SUSY parameter space. In several SUSY breaking scenarios it is hard to 
reconcile Yukawa coupling unification with experimental constraints from $B(b \rightarrow s \gamma )$ and the muon 
anomalous magnetic moment $a_\mu$. We show that $b - \tau$ or even $t - b - \tau$ Yukawa unification can be satisfied 
simultaneously with $b \rightarrow s \gamma $ and  $a_\mu$ in the non-universal gaugino mediation scenario.   
Non-universal gaugino masses naturally appear in higher dimensional grand unified 
models in which gauge symmetry is broken by orbifold compactification.  
Relations between SUSY contributions to fermion masses, $b \rightarrow s
\gamma $ and $a_\mu$ which are typical for models with universal gaugino masses are relaxed. 
Consequently, these phenomenological 
constraints can be satisfied simultaneously with a relatively light SUSY spectrum, compared to models with universal
gaugino masses.
}


\begin{document}

\section{Introduction}
\label{sec:introduction}

Supersymmetric grand unified theories [SUSY GUTs] are well motivated possibilities for physics beyond the
standard model [SM]. They provide an explanation of charge assignments of quarks and leptons~\cite{su4xsu2xsu2, 
su5, 
so10}. Furthermore
the
unification of gauge
couplings in the supersymmetric version of the standard model supports the idea of an underlying theory which
unifies three seemingly unrelated
gauge symmetries at a scale $M_G \sim 2 \times 10^{16}$ GeV~\cite{eft, susygut,
gutexp}.

There is also a possibility that Yukawa couplings of quarks and leptons unify 
in a similar way~\cite{nothreshcorr}. This is certainly not a necessity since 
GUT symmetry breaking effects can be incorporated into Yukawa matrices at the 
GUT scale. 
Furthermore, each of the MSSM Higgs doublets may originate from more than one 
unified Higgs representation in which case the relations between Yukawa couplings can be 
basically arbitrary. (Although given by the underlying theory, these relations might be 
hard to recover from measurements conducted at low energies.)
Nevertheless the hope is that
the underlying theory is quite elegant and
certain simple relations between Yukawa couplings hold. The simplest and the best motivated relation is the
third generation Yukawa coupling unification at the GUT scale: 
$\lambda_b = \lambda_\tau$ in SU(5) or $\lambda_t = \lambda_b = \lambda_\tau$
in SO(10).

A precise analysis of gauge and Yukawa coupling unification requires two loop renormalization group running and 
one loop weak scale SUSY threshold corrections. While gauge coupling unification is not very sensitive to the 
exact form of SUSY spectrum, the fate of Yukawa coupling unification crucially depends on SUSY threshold 
corrections~\cite{threshcorr}. 
This is due to the fact that gluino and chargino corrections to the bottom quark mass are enhanced by $\tan \beta$ 
and can naturally be as large as $ 50 \%$. Therefore the success of Yukawa coupling unification strongly 
depends on the SUSY breaking scenario under consideration. Alternatively, looking at it from the other side, requiring Yukawa 
coupling unification can point to a preferred SUSY breaking scenario or to a region of the parameter space 
within each scenario. 

Similarly $b \rightarrow s \gamma$ and the anomalous
magnetic moment of the muon $a_\mu$ receive contributions from SUSY loops. The
size of SUSY contributions and their signs depend on the SUSY spectrum.
Whether SUSY contributions enhance or suppress these observables compared to the standard model predictions 
mostly depends on relative signs of the gaugino masses $M_1$, $M_2$, $M_3$ and the $\mu$ term.  

For example, in the mSUGRA scenario where all gauginos have the same mass $M_{1/2}$ at $M_{GUT}$, 
Yukawa coupling unification can be satisfied with negative SUSY threshold corrections to the bottom 
quark mass. The gluino--sbottom loop is the dominant contribution and is negative for $\mu M_3 < 0$. 
On the other hand, the chargino contribution to $b \rightarrow s \gamma$ 
can interfere constructively or destructively with the SM
and the charged Higgs contribution. Since the SM and charged Higgs contribution is already somewhat too large, the 
chargino contribution is expected to lower the branching fraction $B (b \rightarrow s \gamma )$. This happens for $\mu A_t <0$, where 
$A_t$ is the top trilinear coupling. The low energy value of $A_t$ is related to $-M_3$ due to renormalization group [RG] 
running. Therefore, in order to accommodate the $b \rightarrow s \gamma$ data, $\mu M_3 >0$ is 
preferred.~\footnote{Constructive interference of the chargino contribution with the SM and charged Higgs contribution 
is not ruled out. However in this case this contribution has to be very small which requires very heavy 
superpartners. This is especially true when considering Yukawa unification, since chargino contribution is 
strongly enhanced in large $\tan \beta$ regime. Note however, for the same reason even the case with preferred 
sign of $\mu A_t$ is strongly constrained by $b \rightarrow s \gamma$.}  
Finally, the SUSY contribution to the anomalous magnetic moment of muon, $a_\mu$, is preferred 
to be positive to comply with the data~\cite{muon_g-2}. The chargino--sneutrino 
loop is dominant in this case and this contribution is positive for $\mu M_2 > 0 $.  
In conclusion, in the mSUGRA scenario both $b \rightarrow s \gamma$ and $a_\mu$ seem to prefer $\mu M_{1/2} 
>0$ while Yukawa coupling unification strongly prefer $\mu M_{1/2} < 0$.
This discouraging result led to consideration of Yukawa coupling unification within other well motivated 
frameworks~\cite{bdr, baer_so10, tobe_wells, yukawa_nonuniversalities, yukawa_other}.

In Refs.~\cite{bdr, baer_so10, Auto:2003ys} SO(10) Yukawa unification was considered together with SO(10) motivated 
boundary conditions for soft SUSY breaking parameters. It was found that $t - b - \tau$ Yukawa unification 
can be satisfied with positive $\mu$ in a special region where the condition $A_0^2 = 4 m_{16}^2 = 2 
m_{10}^2 $ between universal trilinear coupling ($A_0$), universal squark and slepton masses ($m_{16}$) and 
universal Higgs masses ($m_{10}$) is approximately satisfied. 
Additional splitting of Higgs masses 
($m_{H_u}$, $m_{H_d}$) at the level of $10\%$ is necessary in order to obtain electroweak symmetry breaking 
radiatively in large 
$\tan \beta$ regime and large $m_{16}$ ($> 1$ TeV). 
The reason for Yukawa coupling unification to work 
in this region is that chargino corrections to the bottom quark mass are enhanced and can dominate the gluino 
correction, leading to small or even negative total SUSY threshold correction to the bottom quark mass~\cite{bdr}.
This region also has other very compelling features. The masses of first two generation scalars are large, the 
order of $m_{16}$, which suppresses flavor and CP violation (and also proton decay), while the masses of the 
third 
generation squarks and 
sleptons are below 1 TeV keeping this region natural with respect to electroweak symmetry breaking.
 Although this looks like a drastic departure from 
mSUGRA scenario, it actually may originate from it when additional RG running above the GUT scale and GUT scale 
threshold corrections are properly taken into account~\cite{bdr}. 

In anomaly mediation the gaugino masses $M_2$ and $M_3$ have opposite signs and so it might be possible to 
simultaneously achieve 
negative SUSY threshold correction to the bottom quark mass  and positive SUSY contribution to $a_\mu$. However in 
this case the chargino contribution to $b \rightarrow s \gamma$ constructively add to the SM and charged Higgs 
contribution and thus have to be very suppressed. This is possible only if scalar masses are very heavy 
(at least few TeV), which is problematic with respect to naturalness constraints~\cite{tobe_wells}. 

Another approach to accommodate both $b - \tau$ Yukawa coupling unification and constraints from $b \rightarrow 
s \gamma$ and $a_\mu$ was taken up in Ref.~\cite{yukawa_nonuniversalities} where non-universalities in gaugino 
masses in supergravity models were considered. It was found that Yukawa unification can be achieved with an accuracy of a few 
percent 
in significant regions of SUSY parameter space. The corresponding sparticle spectrum is relatively light, 
consistent with naturalness constraints.

Non-universalities in gaugino masses can be very easily obtained (and are quite generic) in higher 
dimensional GUT models in which GUT 
symmetry breaking is achieved by orbifold compactification of extra dimensions. The doublet triplet splitting 
of Higgs fields is also achieved in an elegant way and proton decay due to dimension 5 operators (which is a
serious problem in 4-dimensional GUT models \cite{proton}), can be naturally suppressed in these models.  
The common feature of these models is the existence of a
brane or several branes at orbifold fixed points on which the
gauge symmetry is restricted to be a subgroup of the GUT symmetry.
This results in an effective 4-dimensional theory with  
gauge symmetry given by an intersection of gauge
symmetries on orbifold fixed points~\cite{kawamura, hall_nomura, other_models, so10in5d, so10in6d}. 

The existence of a brane with restricted gauge symmetry 
plays an important role in gaugino mediation.   
If SUSY is broken on a brane with restricted gauge symmetry,
non-universal gaugino masses are generated. For example,
if using proper boundary conditions
$SU(5)$ is broken on a brane down to the SM,
non-universal gaugino masses $M_1, M_2, M_3$ can be
generated on this brane~\cite{hall_nomura}. 
Even more interesting is the situation for $SO(10)$ models in
higher dimensions~\cite{so10in5d, so10in6d} which can
contain branes with gauge symmetries being different subgroups of $SO(10)$.
Using proper boundary conditions, fixed
branes with Pati-Salam $SU(4) \times SU(2)_L \times SU(2)_R$,
Georgi-Glashow $SU(5) \times U(1)$, or
flipped $SU(5)^\prime \times U(1)^\prime$ gauge symmetries can be obtained.
If gauginos get masses on these branes, the
gauge symmetry relates gaugino masses of the MSSM at the
compactification scale. In the case of Pati-Salam gauge symmetry,
gaugino masses $M_2$ and $M_3$ are free parameters while the
$M_1$ is given by a linear combination
of $M_2$ and $M_3$~\cite{so10in5d}. The case of Georgi-Glashow gauge symmetry leads to universal gaugino 
masses 
and finally in the case of flipped $SU(5)^\prime \times U(1)^\prime$ gauge symmetry, $M_2 = M_3$ and $M_1$ is 
an independent parameter~\cite{so10in5d_fsu5}.   
If matter fields are localized on a brane with GUT symmetry Yukawa coupling unification is expected as 
in four dimensional models.

The compactification scale $M_c$ in these models is below (but close to) the GUT scale
and  the boundary conditions with negligible
sfermion masses and trilinear couplings are realized at this scale.
Scalar masses and trilinear couplings receive large
contributions from gaugino masses through the renormalization group (RG)
running between $M_c$ and the electroweak (EW) scale.
These contributions are flavor blind and therefore the
resulting soft SUSY breaking terms at the EW scale
cause only a modest flavor violation originating from the Yukawa couplings.~\footnote{For discussion of other 
SUSY breaking scenarios in higher dimensional GUT models, see for example Ref.~\cite{other_susy_br}.}

In the original works on gaugino mediation~\cite{ginoMSB1, ginoMSB2},
the compactification scale was assumed to be at or above
the GUT scale,
in order to preserve the success of gauge coupling unification.
Therefore all gaugino masses are equal at $M_c$.
For $M_c = M_{GUT}$, however, this scenario predicts
the lightest stau $\ttau_1$ to be the lightest supersymmetric
particle (LSP), which violates cosmological bounds on the existence
of stable charged or colored relics from the Big Bang in models
which conserve $R$-parity.
The situation is different for $M_c > M_{GUT}$.
In this case additional RG evolution 
takes place between $M_c$ and $M_{GUT}$~\cite{ginoMSB2}. This running
generates non-vanishing scalar masses and trilinear couplings at the GUT scale.
Most importantly, the stau mass receives a positive
contribution which eventually can make the $\ttau_1$ heavier than the
lightest neutralino $\tz_1$. This removes the unpleasant
charged LSP feature of the scenario with
$M_c = M_{GUT}~$\cite{ginoMSB2,gaugino_Baer}.

This cure of the
stau LSP problem doesn't apply to the case of non-universal case (in which $M_c < M_{GUT}$). However, in this 
case the
non-universal gaugino masses help us to obtain viable SUSY  
spectra with a neutralino LSP, at least in some 
regions of model parameter space. A recent study~\cite{nonuniversal_gaugino} of the non-universal gaugino 
mediation scenario delineates the allowed regions of the SUSY parameter space consistent with neutralino LSP 
and constraints from $b \rightarrow s \gamma$ and $a_\mu$ for various boundary conditions on gaugino 
masses.

In this paper we study to which extent Yukawa coupling unification can be satisfied together with 
phenomenological constraints from  $b \rightarrow s \gamma$ and $a_\mu$ in the non-universal gaugino mediation scenario. In 
Sec.~\ref{sec:spectrum}
we present basic results of non-universal gaugino mediation and obtain approximate formulas for gaugino,
squark and slepton masses. In Sec.~\ref{sec:thresh_corr} we study SUSY threshold corrections to the bottom quark mass 
based on the SUSY spectrum. Understanding of threshold corrections helps us to understand numerical 
results presented in Sec.~\ref{sec:results_sm}. The summary of our results and our conclusions are given in 
Sec.~\ref{sec:conclusions}.

\section{SUSY spectrum of non-universal gaugino mediation}
\label{sec:spectrum}

Gaugino mediated SUSY breaking is quite economical. In the non-universal 
scenario this mechanism is parametrized by three soft SUSY breaking gaugino 
masses at the compactification scale $M_c$: $M_a (M_c)$, $a = 1,2,3$. 
Depending on the 
localization of the Higgs fields, soft SUSY breaking Higgs masses and the 
$\mu$--term can also be generated. Soft SUSY breaking masses of squarks and sleptons 
and trilinear couplings are negligible at $M_c$. More details on possible models are 
presented in Ref.~\cite{nonuniversal_gaugino} and references therein.

It is an useful exercise to obtain approximate analytic formulas of the SUSY spectrum. These will help 
us in understanding of SUSY threshold corrections to the bottom quark mass and thus the region of SUSY parameter space 
where 
Yukawa coupling unification can be satisfied. 
The complete set of two loop MSSM RG evolution equations [RGEs] can be find in~\cite{RGEs}. 
In this section, 
we use one loop RGEs for the gauge couplings, the gaugino masses and the scalar masses. We 
neglect the contribution of scalar masses and trilinear couplings in the running. 

From the one loop RGEs for gaugino masses and gauge couplings,
\begin{eqnarray}
\frac{d M_a }{d \log \mu} &=& 2 C_a g_a^2 M_a, \quad  C_a = \frac{1}{16 \pi^2} (33/5, 1, -3), \quad a = 
1,2,3  \\
\frac{d g_a }{d \log \mu} &=& C_a g_a^3 ,   
\end{eqnarray}
we obtain the well known result that gaugino masses scale with the square of gauge couplings:
\begin{equation}
\frac{M_a(\mu)}{g_a^2(\mu)} \; = \;  const \; = \; \frac{M_a(M_c)}{g^2},
\end{equation}
where $g$ is the gauge coupling constant at $M_c$. Using weak scale
values: $\alpha_3(M_Z) = 0.118$, $\alpha_{em}(M_Z) = 1/128$,
$\sin^2 \theta_W = 0.22$,
$\alpha_2 = \alpha_{em} / \sin^2 \theta_W \sim 0.035$, $\alpha_1 =
(5/3) \alpha_{em}/ cos^2 \theta_W \sim 0.017$, and the GUT scale value   
of the gauge coupling, $\alpha_G = g^2 / 4 \pi \sim 0.04$, it is easy to find weak scale values of gaugino masses: 
\begin{eqnarray}
M_1 (M_Z) &\sim& 0.4 \, M_1,  \label{eq:M1} \\
M_2 (M_Z) &\sim& 0.9 \, M_2,  \\
M_3 (M_Z) &\sim& 3.0 \, M_3. 
\end{eqnarray}
Similarly, one loop RGEs of squarks and sleptons are given by 
\begin{equation}
\frac{d m_{\tilde f}^2 }{d \log \mu} \; = \; \frac{1}{16 \pi^2} \,
\beta_{m^2_{\tilde f}},
\end{equation}
with
\begin{eqnarray}
\beta_{m^2_Q} &=& - \frac{32}{3} g_3^2 M_3^2 \, - \, 6 g_2^2
M_2^2 \, - \,\frac{2}{15}  g_1^2 M_1^2 \, + \, \dots ,
\nonumber \\
\beta_{m^2_U} &=& - \frac{32}{3} g_3^2 M_3^2
\, - \, \frac{32}{15} g_1^2 M_1^2 \, + \, \dots ,
\nonumber \\
\beta_{m^2_D} &=& - \frac{32}{3} g_3^2 M_3^2
\, - \, \frac{8}{15} g_1^2 M_1^2 \, + \, \dots ,
\nonumber \\
\beta_{m^2_L} &=& - 6 g_2^2 M_2^2
\, - \, \frac{6}{5} g_1^2 M_1^2 \, + \, \dots ,
\nonumber \\
\beta_{m^2_E} &=&
 \, - \, \frac{24}{5} g_1^2 M_1^2 \, + \, \dots ,
\nonumber
\end{eqnarray}
where dots represent terms which are zero at $M_c$, being proportional to scalar 
masses and trilinear couplings. For comparable values of gaugino masses we can neglect contributions from 
terms proportional to $M_1$ (except in the case of right handed sleptons) since these are suppressed 
compared to terms proportional to $M_2$ and $M_3$
by smaller value of gauge coupling $g_1$ and also 
the group theoretical factors. 

In this approximation the solution can be written as:
\begin{eqnarray}
m^2_Q (\mu) &=& \frac{8}{9}  M_3^2  \left[ \left(
\frac{g_3 (\mu)}{g} \right)^4 - 1 \right] \, + \, \frac{3}{2} 
M_2^2  \left[ 1 - \left( \frac{g_2 (\mu)}{g} \right)^4
\right],  \\
m^2_U (\mu) &=& \frac{8}{9}  M_3^2  \left[ \left(
\frac{g_3 (\mu)}{g} \right)^4 - 1 \right],  \\
m^2_D (\mu) &=& \frac{8}{9}  M_3^2  \left[ \left(
\frac{g_3 (\mu)}{g} \right)^4 - 1 \right],  \\
m^2_L (\mu) &=& \frac{3}{2} 
M_2^2  \left[ 1 - \left( \frac{g_2 (\mu)}{g} \right)^4
\right],  \\
m^2_E (\mu) &=& \frac{2}{11} 
M_1^2  \left[ 1 - \left( \frac{g_1 (\mu)}{g} \right)^4
\right], 
\end{eqnarray}
and for the weak scale values we obtain:
\begin{eqnarray}
m_Q (M_Z) &=& 2.6 \, M_3 \, \left[ 1 + 0.03
\left( \frac{ M_2  }{ M_3 }
\right)^2 \right],  \\
m_U (M_Z) &=& m_D (M_Z) \; = \; 2.6 \, M_3 , \\
m_L (M_Z) &=& 0.6 \,  M_2 , \\
m_E (M_Z) &=& 0.39 \,  M_1 \label{eq:mE} .
\end{eqnarray}

We also need to estimate the top trilinear coupling which enters the chargino contribution to the bottom quark mass. 
The one loop RGE is given by 
\begin{equation}
\frac{d A_t }{d \log \mu} \; = \; \frac{1}{16 \pi^2} \, \beta_{A_t} ,
\end{equation}
where
\begin{equation}
\beta_{A_t} =  \frac{32}{3} g_3^2 M_3 \, + \, 6 g_2^2
M_2 \, + \,\frac{26}{15}  g_1^2 M_1 \, + \, \dots \, .
\end{equation}
In the same approximation as for the squarks and sleptons, we obtain
\begin{equation}
A_t (\mu) =  \frac{16}{9}  M_3  \left[ 1 - \left(
\frac{g_3 (\mu)}{g} \right)^2 \right] \, - \, 3 
M_2  \left[ 1 - \left( \frac{g_2 (\mu)}{g} \right)^2  
\right], 
\end{equation}
and the weak scale value is given by
\begin{equation}
A_t (M_Z) = - 3.5 \,  M_3  \, - 0.4 M_2 .
\end{equation}


\FIGURE[t]{
\epsfig{file=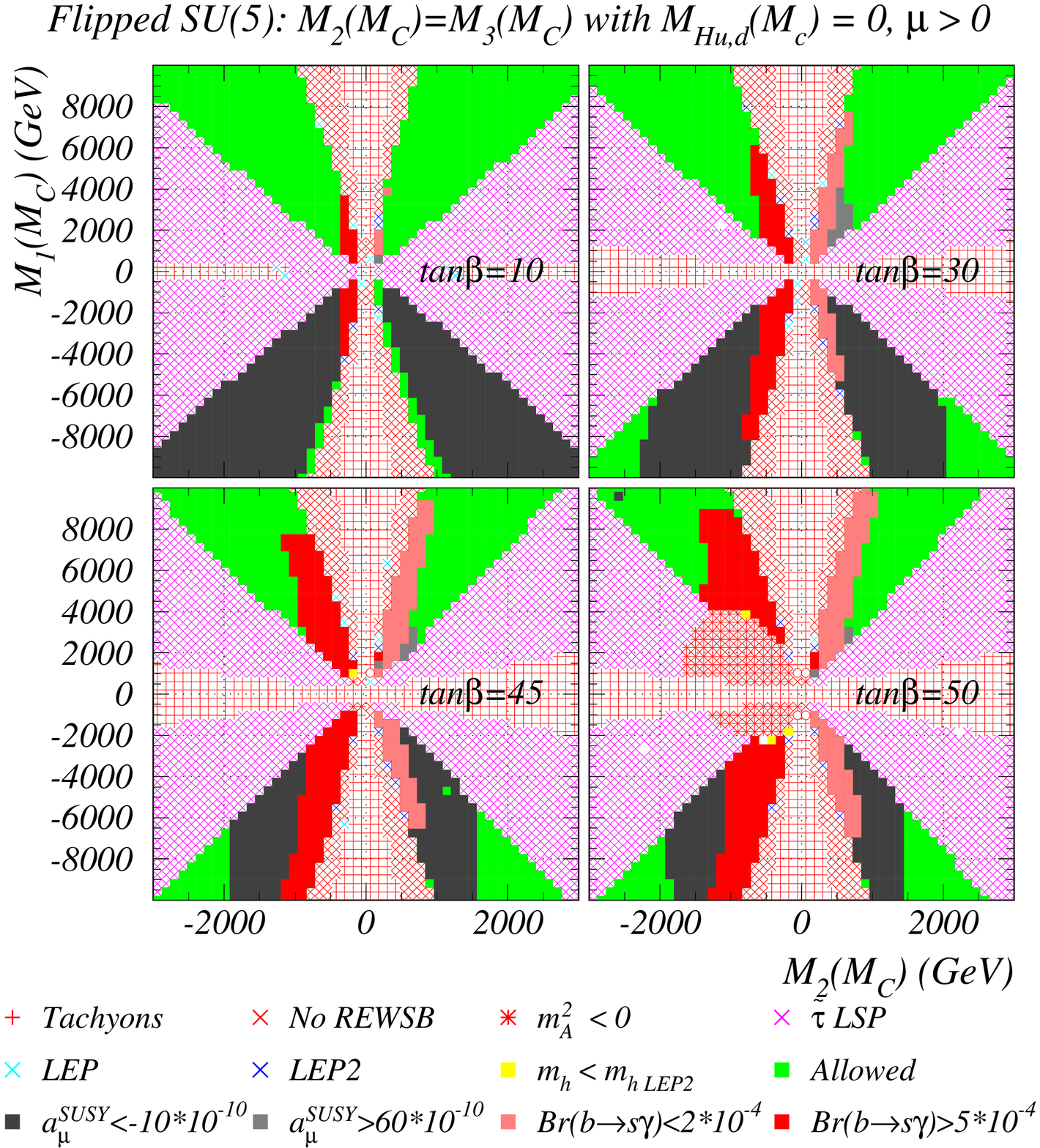,height=11cm}
\caption{
Typical parameter space slices of flipped SU(5) gaugino mediation models. Areas
hatched in red are excluded by the presence of tachyonic particles or by the
lack of radiative EW symmetry breaking. The magenta areas represent regions with
stau LSPs. Blue and yellow areas are excluded by LEP.
}
\label{fig:planeFSU5}}

These results for gaugino, squark and slepton masses should be interpreted  with care. However the approximation 
used is precise enough  
to point out the main features of the SUSY spectrum in non-universal gaugino mediation. First of all, from 
Eqs.~(\ref{eq:M1}) and (\ref{eq:mE})  we 
immediately see the well-known problem associated with gaugino mediation, the stau being LSP. Neither the bino nor the 
right-handed stau are mass eigenstates however. The lightest neutralino is a mixture of gauginos and Higgsinos, and 
for $\tilde \tau_1$ the D-terms as well as left-right mixing have to be properly taken into account. In result, 
there is quite significant region of parameter space with neutralino LSP, 
as illustrated by Fig.~\ref{fig:planeFSU5}. 
The allowed region opens up with increasing $M_1$, either because of the large portion of Higgsino in the lightest 
neutralino (along the line where electroweak symmetry breaking is possible, where the $\mu$ term is small) or the large 
portion 
of wino in the lightest neutralino (for small values of $M_2$ compared to $M_1$). On the other hand 
with increasing $\tan \beta$ the $\tau$ Yukawa coupling also increases and the terms proportional to Yukawa 
couplings in RG equations become 
more important. These terms suppress the 
stau mass and so the allowed region is shrinking. For more results and discussion 
see Ref.~\cite{nonuniversal_gaugino}.

The second important observation is that the gluino, stop and sbottom masses are very close to each other and 
approximately equal to $3 M_3$. This will be crucial for discussion of SUSY threshold corrections to the bottom 
quark mass in the following section.


\section{SUSY threshold corrections to the bottom quark mass}  
\label{sec:thresh_corr}

The dominant corrections to the bottom quark mass come from gluino-sbottom and chargino-stop diagrams. Pieces 
proportional to 
$\tan \beta$ can be approximated by: 
\begin{equation}
\left( \frac{\delta m_b}{m_b} \right)^{\tilde g} 
\sim \; \frac{2 \alpha_3}{3 \pi} \;
\mu \tan \beta \; \frac{m_{\tilde g}}{M_{SUSY}^2} 
\end{equation}
and 
\begin{equation} 
\left( \frac{\delta m_b}{m_b} \right)^{\tilde \chi^\pm}
 \sim \; \frac{\lambda^2_t}{16 \pi^2} A_t \, \tan \beta
\; \frac{\mu}{M_{SUSY}^2} .
\end{equation}
Here $M_{SUSY}$ is the typical mass of particles in the corresponding loop.  
In the previous section we found that
\begin{eqnarray}
M_{SUSY}  & \sim &  m_{\tilde g} \, \sim \,  m_{\tilde b} \, \sim \, m_{\tilde t} \, \sim \, 3 M_3 , \\
A_t & \sim & - 3.5  M_3  \, - 0.4 M_2 .
\nonumber
\end{eqnarray}
Taking $\alpha_3 (M_Z) = 0.118$ and $\lambda_t (M_Z) \sim 1$,
we can obtain very simple formulas for the SUSY threshold corrections to $m_b$:
\begin{eqnarray}
 \left( \frac{\delta m_b}{m_b} \right)^{\tilde g}  
& \sim & \frac{1}{100  M_3 } \; \mu \tan \beta , \\
 \left( \frac{\delta m_b}{m_b} \right)^{\tilde \chi^\pm}
 & \sim & - \frac{1}{500
M_3 } \left( 1 + \frac{1}{9} \frac{ M_2
 }{M_3  } \right) \mu \tan \beta .
\end{eqnarray}   
Comparing the last two expressions we find
\begin{equation}
 \left( \frac{\delta m_b}{m_b} \right)^{\tilde \chi^\pm}
\sim  - \frac{1}{5}
\left( \frac{\delta m_b}{m_b} \right)^{\tilde g}
\left( 1 + \frac{1}{9} \frac{ M_2 }{ M_3  } \right).
\end{equation}

Clearly, the gluino corrections will dominate unless $M_2 \sim 40 M_3$. 
So for positive $\mu$, the Yukawa unification can be expected only
for negative $M_3$, and the dependance on $M_2$ and $M_1$ is not expected to be significant.
The results shown in the figures of the next section are in agreement with these findings.
Although $\mu M_3 <0$ is strongly preferred as in the mSUGRA scenario, non-universal gaugino masses can help 
to satisfy constraints from $b \rightarrow s \gamma$ and $a_\mu$ in some portion of SUSY parameter space.


\section{Yukawa coupling unification}
\label{sec:results_sm}

In this section, we present results of our numerical analysis. There are several 
programs readily available for numerically solving the necessary, coupled system 
of RGEs at the two-loop level. The most popular ones are ISAJET \cite{isajet}, 
SoftSUSY \cite{softsusy}, Spheno \cite{spheno} and Suspect \cite{suspect}. 
Recently, a satisfactory agreement was demonstrated between the latest versions 
of these codes \cite{kraml}. To calculate the sparticle spectrum and the Yukawa 
couplings between $M_c$ and the weak scale, we use ISAJET version 7.64. The 
calculational procedure implemented in ISAJET, with special emphasis on Yukawa 
couplings and the theoretical uncertainties in their determination, is described 
in a recent publication \cite{Auto:2003ys}. Due to the known theoretical 
uncertainties, we treat our numerical results with a reasonable flexibility, and 
conclude that unification takes place when the GUT scale Yukawas agree within
a few percent. For the evaluation of the branching fraction $B(b\to 
s\gamma)$ and the muon anomalous magnetic moment $a_\mu$, we use the 
calculations presented in Refs. \cite{Baer:2002gm} and \cite{Baer:2001kn}, 
respectively. 

In our plots, we present theoretically, phenomenologically and 
experimentally acceptable models. We discard models with tachyonic particles, 
no radiative EW symmetry breaking [REWSB] or a stau LSP as unacceptable. 
We further require that the $Z$ boson have 
negligible decay rates to sparticles and also impose the following LEP2 
constraints on the weak scale masses of the lightest gauginos ($\tz_1$, 
$\tw_1$), sleptons ($\te_1$,$\tmu_1$,$\ttau_1$) and Higgs boson \cite{lep2}:
\begin{eqnarray}
&& m_{\tz_1}>37~{\rm GeV}, ~~ m_{\tw_1}>100~{\rm GeV},  \nonumber \\
&& m_{\te_1}>92~{\rm GeV}, ~~ m_{\tmu_1}>85~{\rm GeV}, ~~ 
   m_{\ttau_1}>68~{\rm GeV}, \\ 
&& m_h>91~{\rm GeV}. \nonumber
\end{eqnarray}
Models that do not satisfy the above criteria are excluded from our study. 
Models passing all the above criteria are further analyzed with respect to 
the indirect experimental constraints from $B(b\to s\gamma)$ and $a_\mu$. We impose the following 
limits:
\begin{eqnarray}
-10 < a_{\mu} \times10^{10} < 60, ~~~ {\rm and} ~~~ 
2 < B(b\to s\gamma) \times 10^{4} < 5.
\end{eqnarray}
These limits correspond to the experimentally allowed regions approximately at 
2$\sigma$ \cite{Baer:2002ay}. In our plots, we color models that do not satisfy 
this $B(b\to s\gamma)$ (or $a_\mu$) constraint by shades of red (or gray). 
Models that pass all the above requirements are marked by green on our figures.

As discussed earlier, for gaugino mediation only the gaugino and Higgs masses 
are non-zero at $M_c$. 
In our numerical analysis we assume $M_c = 10^{16}$ GeV.
The rest of the soft breaking parameters, namely all squark and slepton masses 
and trilinear couplings are assumed to vanish at $M_c$. The only 
other remaining free model parameter is $\tan \beta$, since our requirement 
of the REWSB fixes the rest of the Higgs sector. Since we scan through both positive 
and negative values of gaugino masses, the sign of $\mu$ is not an independent 
parameter and we 
present results only for $\mu > 0$.~\footnote{The MSSM Lagrangian is
invariant under the simultaneous sign
change of the gaugino masses, the $A$ and $B$ parameters and
$\mu$. Thus, the results for $\mu <0$ are equivalent to those for $\mu >0$ with gaugino masses
taken with opposite signs, since the trilinear couplings are zero at $M_c$. We thank X. Tata for 
emphasizing this point.} 
Thus, the parameter space of the non-universal gaugino mediation scenario 
is spanned by six parameters:
\begin{equation}
M_1(M_c), ~~ M_2(M_c), ~~ M_3(M_c), ~~ M_{H_u}(M_c), ~~ M_{H_d}(M_c)  ~~ 
{\rm and} ~~ \tan\beta . 
\label{Eq:FullParaSpace}
\end{equation}
This corresponds to the case when gaugino masses are generated on a brane with SM gauge symmetry. 
If SUSY breaking happens on a brane with larger gauge symmetry, gaugino masses are further restricted~\cite{so10in5d}.
In the case Pati-Salam symmetry gaugino masses satisfy
\begin{equation}
M_1={3\over 5} M_2+ {2\over 5}M_3,
\label{eq:PS}
\end{equation}
and in the case of flipped SU(5)
\begin{equation}
M_2 = M_3.
\label{eq:flippedSU(5)}
\end{equation}
In what follows we present results for $b - \tau$ and $t - b - \tau$ Yukawa coupling unification 
in the SM scenario and 
the flipped SU(5) scenario. In both cases we also show results for vanishing Higgs masses, 
$ M_{H_u}(M_c) =  M_{H_d}(M_c) = 0$. Finally, we briefly comment on the Pati-Salam scenario.

\subsection{Bottom - tau Yukawa unification with independent gaugino masses}

\FIGURE[t]{
\epsfig{file=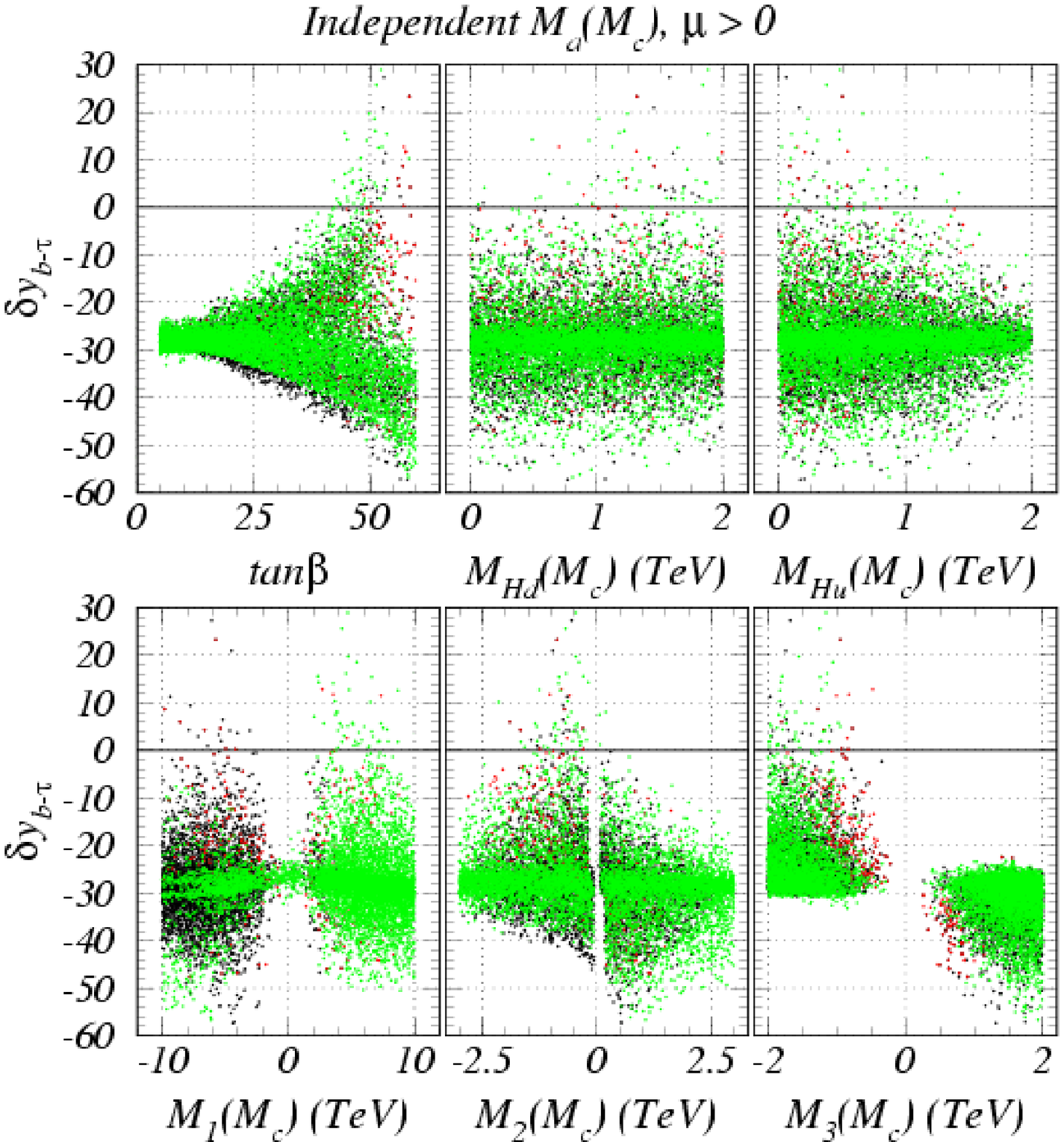,height=11cm}
\caption{
A scan of the full parameter space with independent gaugino and Higgs masses at
$M_c$ and $\mu>0$. The variable $\delta y_{b-\tau}$, defined in Eq.(\ref{Eq:Defybt}), is
plot versus the model parameters. Black (red) dots mark models that deviate from
the central value of the $a_{\mu}$ ($B(b \to s \gamma)$) measurement by
2$\sigma$. Models marked by green dots satisfy all our constraints.
}
\label{fig:scan1mup}}

To quantify the amount of $b-
\tau$ unification, we use the variable
\begin{equation}
\delta y_{b-\tau} = 100 \left( \frac{y_b}{y_\tau} - 1 \right) ,
\label{Eq:Defybt}
\end{equation}
where $y_b$  and $y_\tau$ are the values of the $b$ and $\tau$ 
Yukawa couplings at the compactification scale. The factor 100 is introduced 
so that $\delta y_{b-\tau}$ measures the amount of percentage deviation from perfect
unification.

To find the regions with the best Yukawa unification, namely regions with 
$\delta y_{b-\tau} \sim 0$, we scan the full parameter space as indicated by 
Eq.(\ref{Eq:FullParaSpace}). Fig.~\ref{fig:scan1mup} shows the parameter ranges 
and the result of such a random scan. Here, we plot $\delta y_{b-\tau}$ versus the free 
parameters of the model for $\mu>0$. We observe two branches of models, 
differentiated by the sign of $M_3$. As expected from the inspection of the 
SUSY threshold corrections in Sec.~\ref{sec:thresh_corr}, the branch with negative $M_3$ unifies the $b$ 
and $\tau$ Yukawa couplings while the 
other does not. Yukawa unification happens in relatively wide parameter ranges. 
But models that simultaneously comply with unification and the indirect 
experimental constraints are confined close to $\tan\beta \sim 45$, $M_1(M_c) 
\gtsim 2$ TeV, --2 TeV $\lesssim M_2(M_c) \lesssim 1$ TeV, and $M_3(M_c) 
\lesssim -1$ TeV. In contrast, there is no strong preference for particular 
values of Higgs masses, except perhaps for low $M_{H_u}$ values. This latter 
fact is supported by an independent scan with vanishing Higgs masses 
$M_{h_u}(M_c) = M_{h_d}(M_c) = 0$, shown in Fig.~\ref{fig:scan2mup}. For this 
case $b-\tau$ unification occurs in similar ranges as for non-zero Higgs masses.

\FIGURE[t]{
\epsfig{file=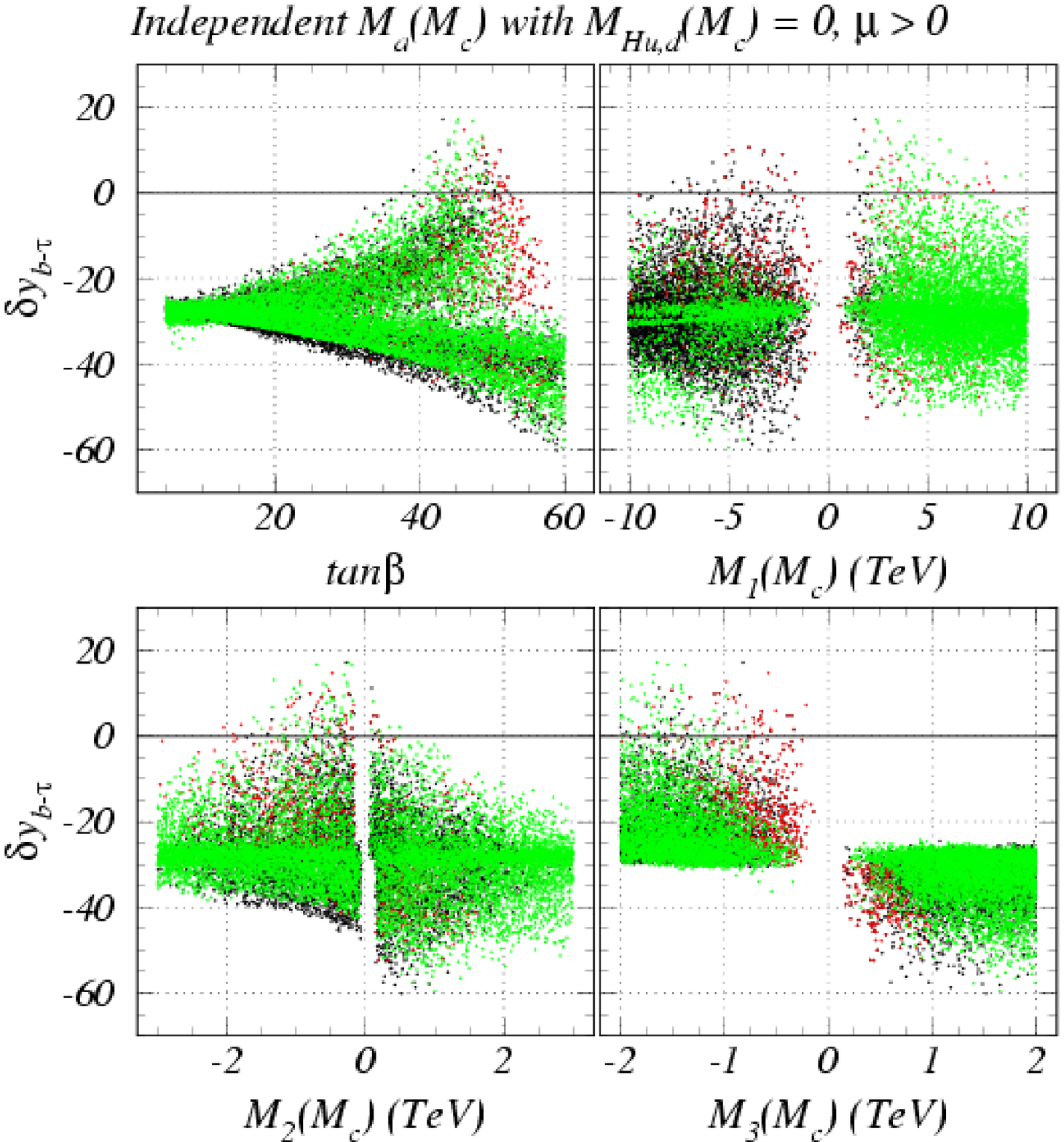,height=11cm}
\caption{
Same as Fig.~\ref{fig:scan1mup} except with vanishing Higgs masses at $M_c$.
}
\label{fig:scan2mup}}




\FIGURE[t]{
\epsfig{file=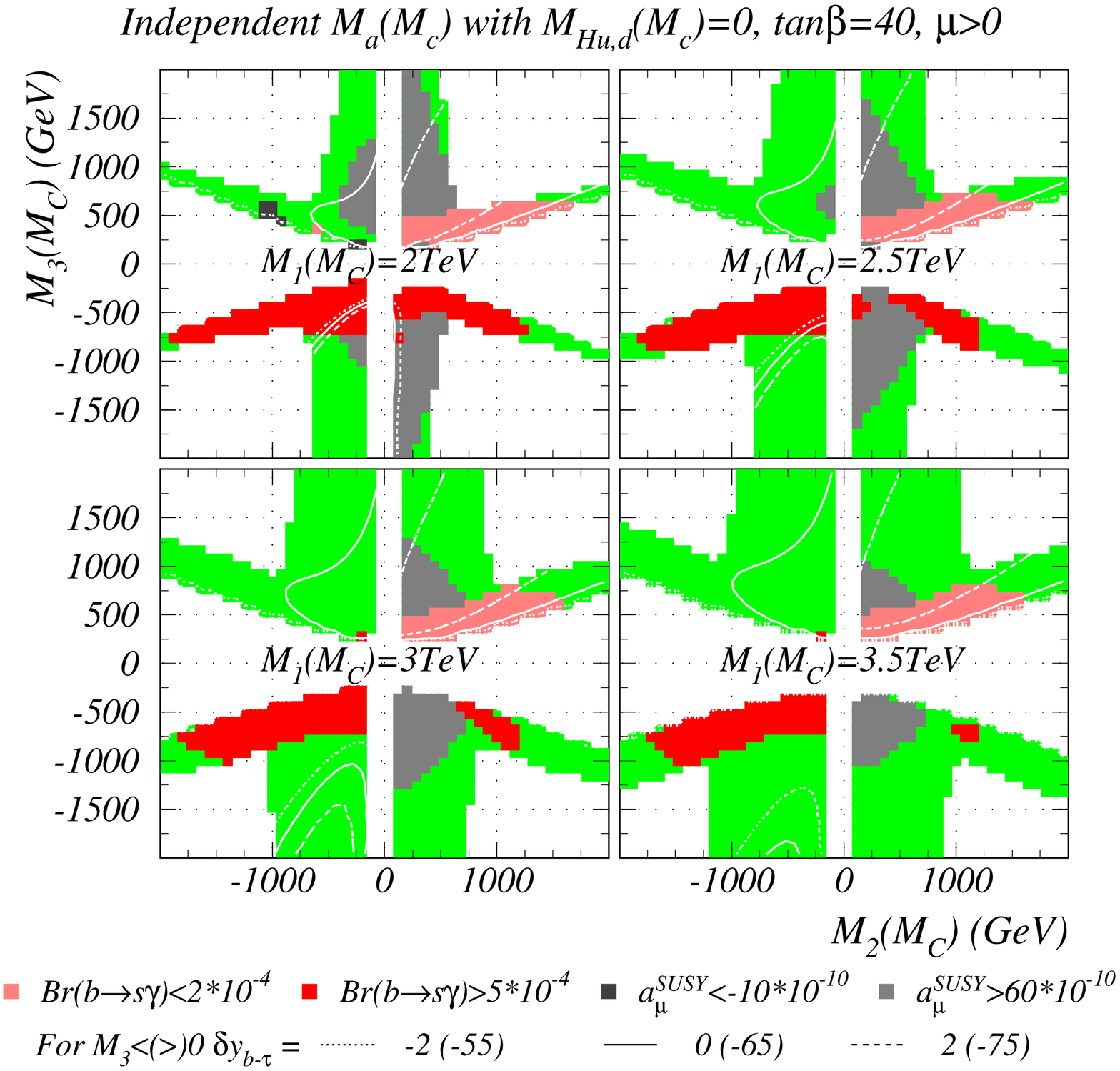,height=11cm}
\caption{
Yukawa unification contours in the $M_3$ versus $M_2$ plane for independent
gaugino and vanishing Higgs masses at $M_c$. We chose $\tan\beta = 40$ and
$\mu > 0$.
}
\label{fig:plane1}}

Figs.~\ref{fig:scan1mup} and \ref{fig:scan2mup} point to specific regions of the 
parameter space where Yukawa unification is achieved and the constraints on $B(b 
\to s \gamma)$ and $a_\mu$ are satisfied simultaneously. 
To 
explore this parameter region further, we fix $\tan\beta = 40$ and scan the 
$M_2$ versus $M_3$ planes for fixed values of $M_1$. Fig.~\ref{fig:plane1} shows 
the results of these scans. White areas are theoretically unacceptable and areas 
colored by shades of red (gray) are excluded by $B(b \to s \gamma)$ ($a_\mu$). 
Green areas are allowed by all the constraints.
As expected, in the negative $M_2$ and $M_3$ quadrants there are large regions 
with good Yukawa unification.  
For $M_1 > 2$ TeV significant parts of these regions are allowed both by
$b\to s\gamma$ and $a_\mu$. Thus, Fig.~\ref{fig:plane1} clearly demonstrates 
that, for the case of independent gaugino and vanishing Higgs masses, there exist
parts of the parameter space where it is 
possible to reconcile $b-\tau$ unification with all the considered constraints.

\subsection{Top - bottom - tau Yukawa unification with independent gaugino masses}

\FIGURE[t]{
\epsfig{file=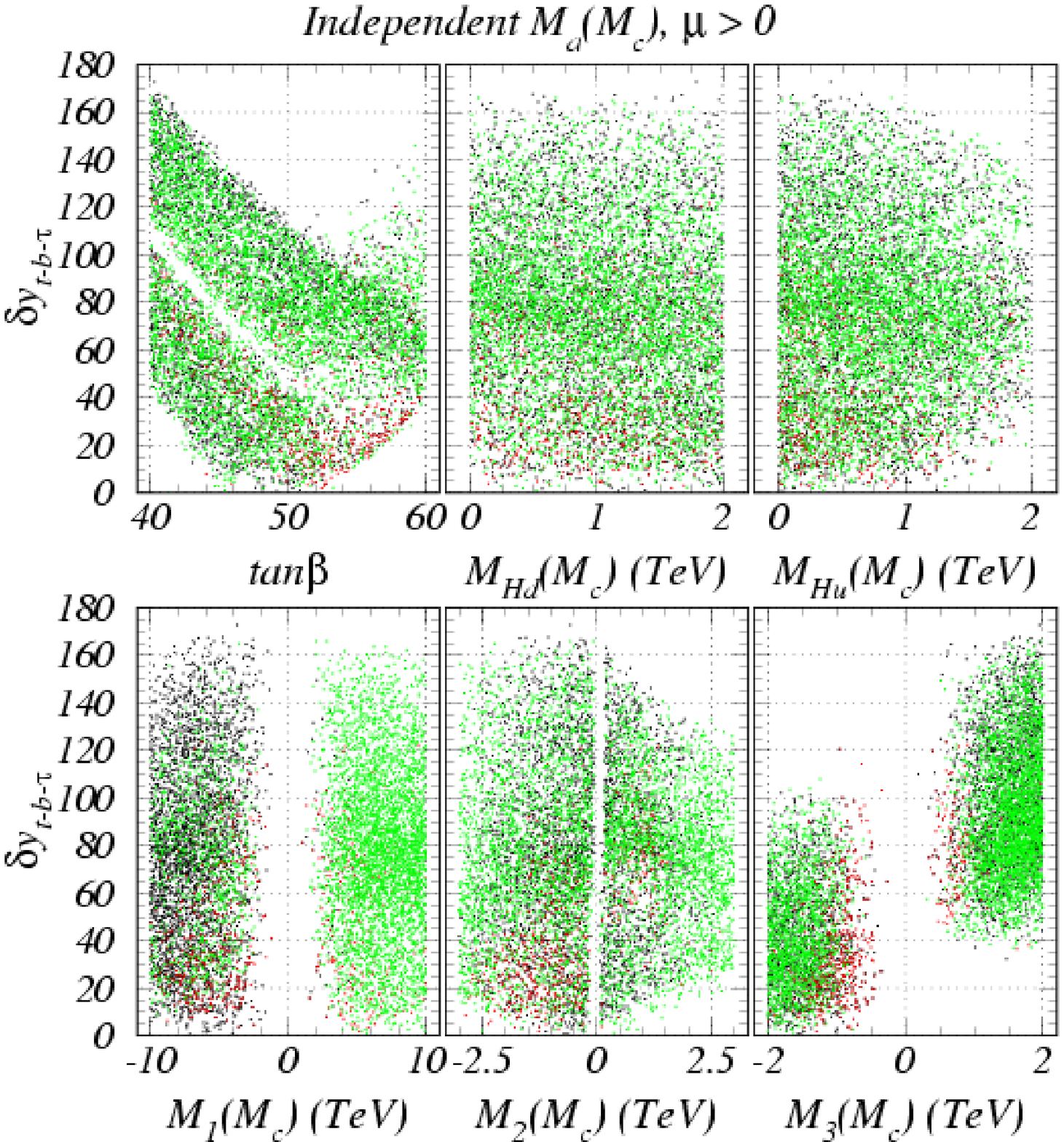,height=11cm}
\caption{
A scan of the relevant parameter space with independent gaugino and Higgs masses 
at $M_c$ and $\mu>0$. The variable $\delta y_{t-b-\tau}$ is plot vs. the model parameters. Black 
(red) dots mark models that deviate from the central value of the $a_{\mu}$ 
($B(b \to s \gamma)$) measurement by 2$\sigma$. Models marked by green dots 
satisfy all our constraints.
}
\label{fig:scan3mup}}

To gauge the amount of $t-b-\tau$ unification, similarly to 
Eq.(\ref{Eq:Defybt}), we define
\begin{equation}
\delta y_{t-b-\tau} = 100 \left( \frac{\max(y_t,y_b,y_\tau)}
                             {\min(y_t,y_b,y_\tau)} 
                        - 1 \right) ,
\end{equation}
where $y_x$ are the values of the $t$, $b$ and $\tau$ Yukawa couplings at $M_c$. 
Notice that unlike $\delta y_{b-\tau}$, $\delta y_{t-b-\tau}$ never becomes negative.
In Fig.~\ref{fig:scan3mup}, we plot $\delta y_{t-b-\tau}$ versus the relevant parameters 
from a scan for $\mu > 0$ in the high $\tan\beta$ region where Yukawa 
unification is expected. Again, we find models with good unification 
satisfying all the constraints. The conclusion is similar to that of the $b-
\tau$ unification case. Provided that we allow high enough $|M_1(M_c)|$ values, 
unification is achieved at moderate values of $M_2$ and $M_3$.
Since the correlation between $\delta y_{t-b-\tau}$ and the Higgs masses seems 
to be weak just like in the case of $\delta y_{b-\tau}$, we conducted a similar 
scan for $M_{h_u}(M_c) = M_{h_d}(M_c) = 0$. The results are shown by 
Fig.~\ref{fig:scan4mup}. In a considerable part of the parameter space, we
find well unified models consistent with all other constraints.

\FIGURE[t]{
\epsfig{file=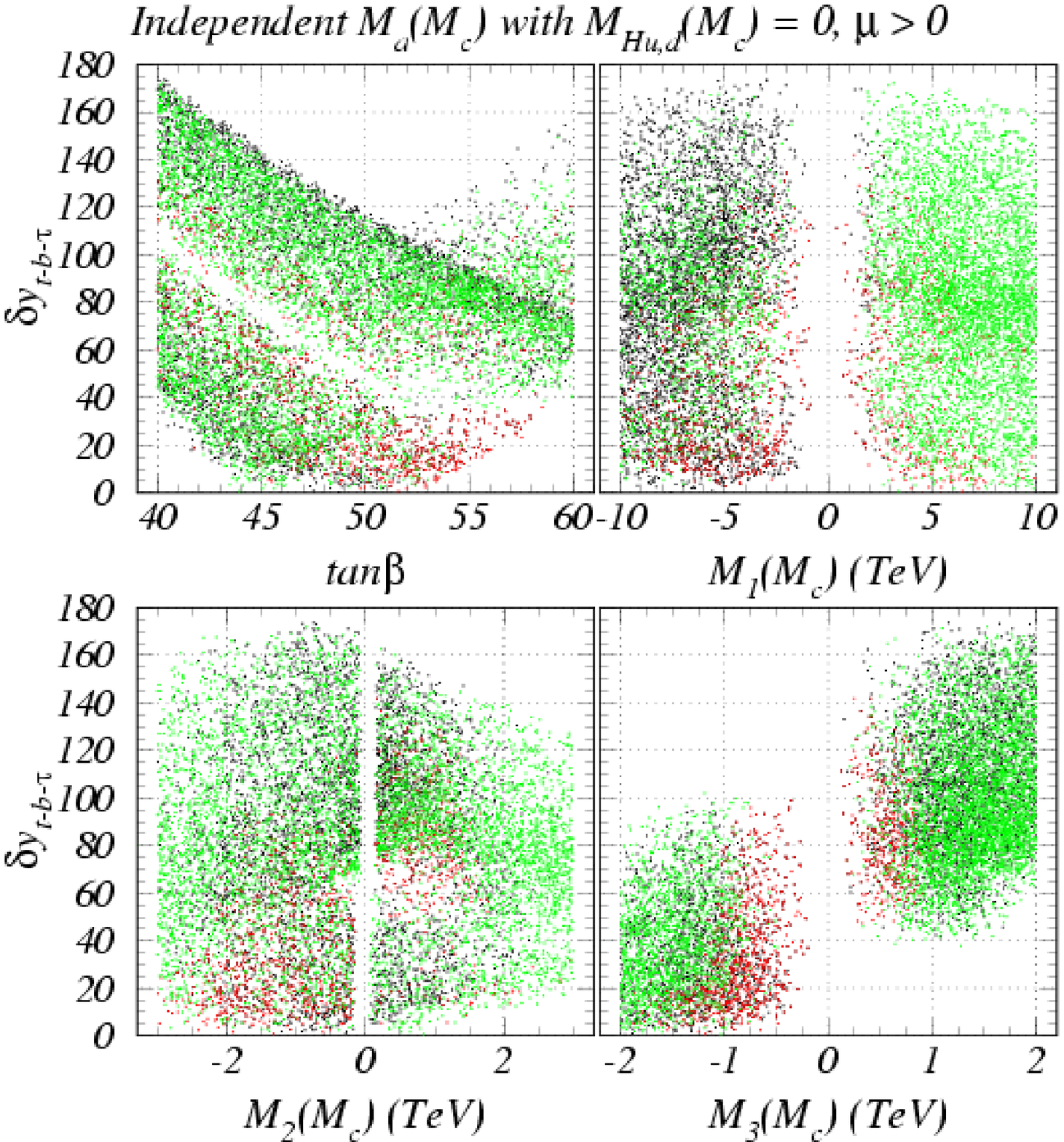,height=11cm}
\caption{
Same as Fig.~\ref{fig:scan3mup} except with vanishing Higgs masses at the
compactification scale.
}
\label{fig:scan4mup}}

Both Fig.~\ref{fig:scan3mup} and Fig.~\ref{fig:scan4mup} show that Yukawa unification 
is achieved for two narrow regions of $\tan \beta$, namely 45 and 50. These two regions 
are correlated with the sign of $M_2$. The region with $\tan \beta \sim 45$ corresponds 
to $M_2 > 0$ while the region with $\tan \beta \sim 50$ corresponds to $M_2 < 0$. 
This can be seen on Fig. \ref{fig:plane2}, where we present contours of $t-b-\tau$
unification down to 5\%. This plot illustrates for $\tan\beta = 45$ that with vanishing 
Higgs masses and $M_1 \sim 5-7$ TeV there are significant regions of the 
$M_2$-$M_3$ parameter plane where $t-b-\tau$ unification takes place. In particular, 
Yukawas unify for $M_2 > 0$. A similar plot, which we do not include here, for 
$\tan\beta = 50$ reveals $t-b-\tau$ unification for $M_2 < 0$.

\FIGURE[t]{
\epsfig{file=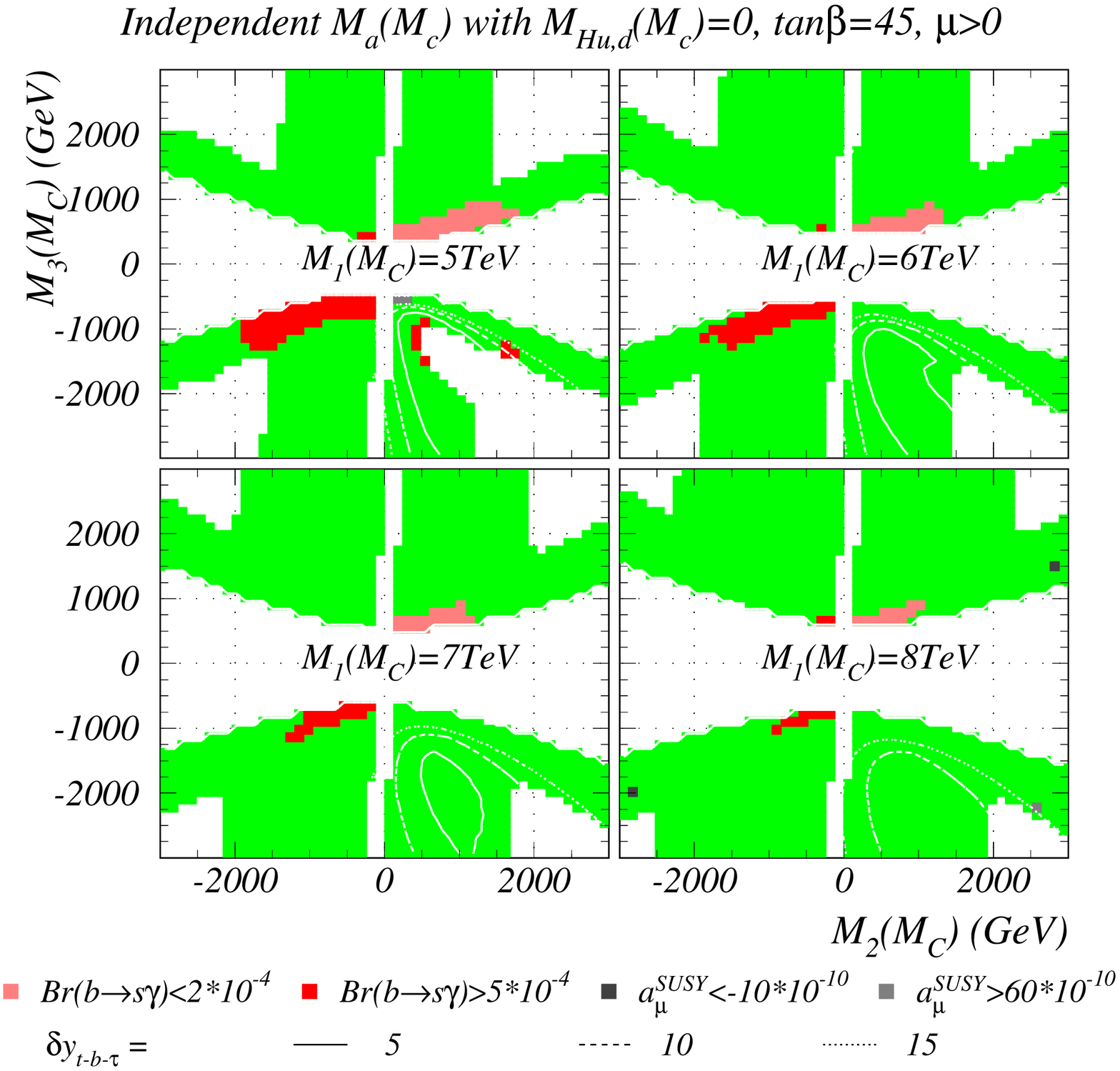,height=11cm}
\caption{
Same as Fig. \ref{fig:plane1} with contours of $t-b-\tau$ Yukawa unification
being shown. On this plot $\tan\beta = 45$.
}
\label{fig:plane2}}

In order to illustrate typical sparticle spectra for models with good unification, 
we show a few parameters for selected models in Table~\ref{tab:one}.
The firs two models have non-vanishing Higgs masses and the third model has zero  
Higgs masses at the compactification scale. We find that well unified
models typically have the lightest gauginos in the few hundred
GeV mass range, lightest sleptons and Higgses below 1 TeV, and the lightest
squarks in the TeV range. This sort of spectrum may allow the Tevatron to
cover a limited part of the parameter space, while the LHC has a chance to
produce several types of sparticles. A LC, depending on its center of mass
energy and the actual model parameters, may produce only the lightest Higgs
boson or additionally the lightest gauginos and sleptons.

As Table~\ref{tab:one} illustrates, the lightest neutralino is always almost degenerate
with the lightest chargino, that is it always has a strong wino admixture. As a
consequence, in this model the neutralino relic abundance is generally much lower
than the experimental value of the cold dark matter \cite{Bennett:2003bz}, as it
was pointed out in \cite{nonuniversal_gaugino}.

\subsection{More constrained cases}

Results for independent gaugino masses indicate that Yukawa unification happens for a wide range of  $M_2$,
which suggests that it can be achieved also 
in the case of gaugino masses being restricted by flipped SU(5) symmetry (cf. Eq.~(\ref{eq:flippedSU(5)})). 
Motivated by this, we re-scan the full parameter space with $M_2 = M_3$ and find that the results are qualitatively 
very similar to those of independent gaugino 
masses, both in the $b - \tau$ and $t - b - \tau$ cases. For completeness, we present results for $t - b - \tau$ Yukawa unification 
in Fig.~\ref{fig:scan_fsu5_mup}. This is a scan with independent Higgs masses although these are not showed in the figure 
since there is no preference for particular values of Higgs masse just as in previous cases. 
As expected, only solutions with $\tan \beta \sim 50$ are found in this case.
We also performed a scan with vanishing 
Higgs masses with similar results. It is interesting to note that even in this case, when the SUSY breaking scenario is 
characterized by just three soft SUSY breaking parameters: $M_1, M_2=M_3$ and $\tan \beta$, we find a region with good Yukawa  
unification as well as $b \rightarrow s \gamma$ and $a_\mu$. In this case Yukawa unification is achieved at slightly higher 
values of 
gaugino masses  compared to the scenario with independent gaugino 
masses: $M_3 \lesssim 1.5$ TeV, $M_1 \gtsim 6$ TeV. Typical SUSY spectra from this region are represented by model 4 in 
Table~\ref{tab:one}. The spectrum in this case exhibits similar features as spectra for independent gaugino masses.

\FIGURE[t]{
\epsfig{file=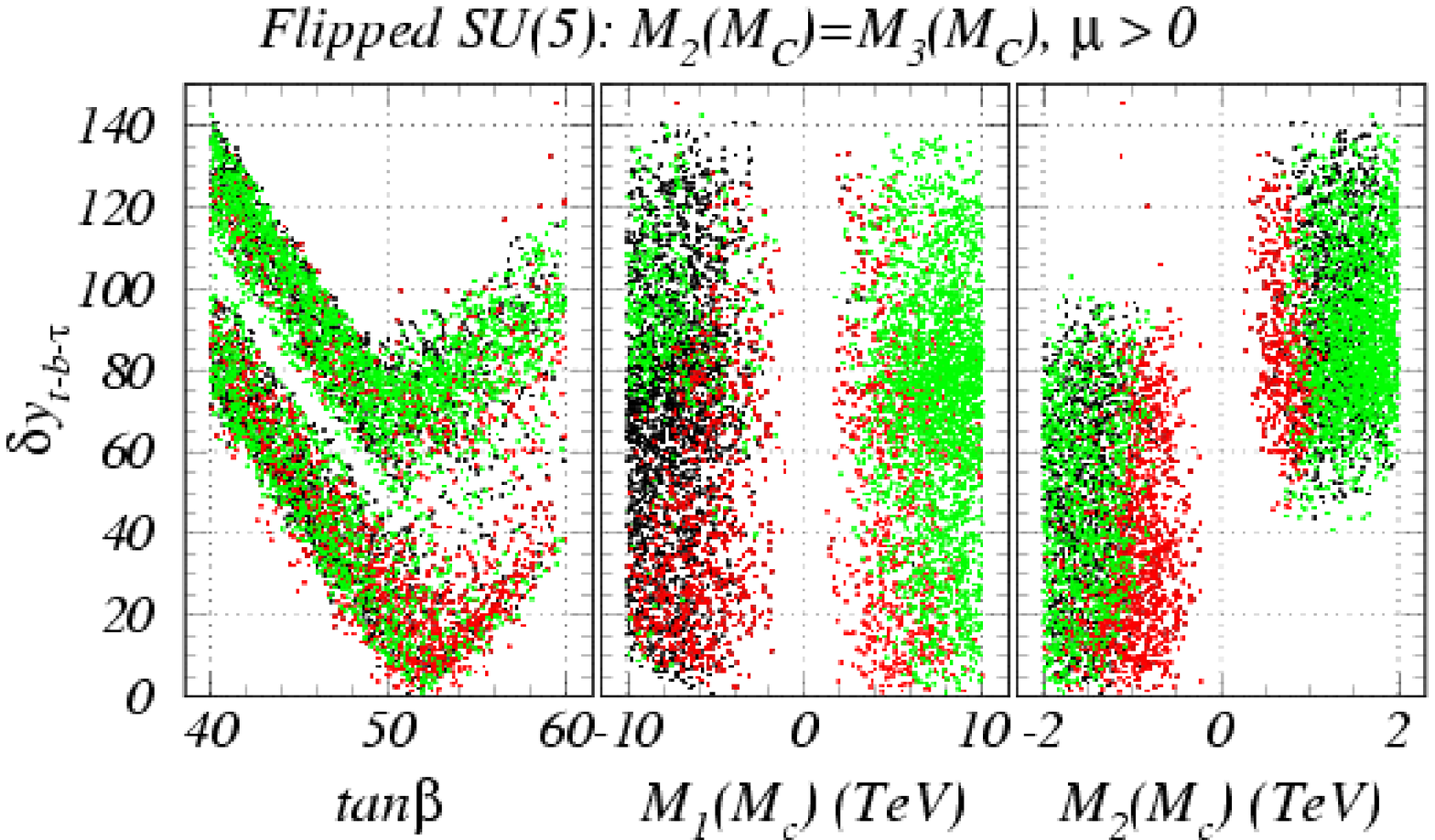,height=6.3cm}
\caption{
A scan of the relevant parameter space with flipped SU(5) related gaugino masses 
at $M_c$ and $\mu>0$. The variable $\delta y_{t-b-\tau}$ is plot vs. the model parameters. Black
(red) dots mark models that deviate from the central value of the $a_{\mu}$
($B(b \to s \gamma)$) measurement by 2$\sigma$. Models marked by green dots
satisfy all our constraints.
}
\label{fig:scan_fsu5_mup}}

Finally, in the case with gaugino masses constrained by Pati-Salam 
symmetry~(\ref{eq:PS}), we do not expect Yukawa unification while keeping 
a relatively light SUSY spectrum.
%
%
Based on results for independent gaugino 
masses and the flipped SU(5) case we see that $|M_1| \gg |M_2|$ is highly preferred. However, this can be 
only achieved for large values of $M_3$. Therefore Yukawa unification in this case might only be satisfied
with gluino and squarks not lighter than several TeV which makes this scenario phenomenologically less interesting.

\TABLE{
\begin{tabular}{lrrrrr}
\hline
model                       &     1 &     2 &     3 &     4 \\
\hline
$M_1(M_c)$                  & 3929.6& 3778.4& 4596.2& 6360.3\\
$M_2(M_c)$                  &  481.3&  777.9&  959.5&-1613.7\\
$M_3(M_c)$                  &-1532.9& -866.9&-1989.5&-1613.7\\
$M_{H_u}(M_c)$              &  950.7&  816.1&    0.0&  996.9\\
$M_{H_d}(M_c)$              &  144.9&   58.8&    0.0&  138.9\\
$\tan\beta$                 &   45.2&   44.7&   44.4&   51.1\\
\hline
$f_t(M_{GUT})$              &  0.525&  0.516&  0.524&  0.509\\
$f_b(M_{GUT})$              &  0.522&  0.510&  0.507&  0.507\\
$f_\tau (M_{GUT})$          &  0.525&  0.532&  0.519&  0.488\\
$\delta_{b-\tau}$           &   -0.6&   -4.1&   -2.2&    4.0\\
$\delta_{t-b-\tau}$         &    0.6&    4.3&    3.3&    4.3\\
$m_{\tg}$                   & 3376.6& 1986.8& 4325.5& 3545.0\\
$m_{\tu_L}$                 & 2837.5& 1738.1& 3647.0& 3104.6\\
$m_{\td_R}$                 & 2864.7& 1721.4& 3642.8& 3044.4\\
$m_{\tst_1}$                & 2500.1& 1494.5& 3231.9& 2649.5\\
$m_{\tb_1}$                 & 2484.0& 1454.2& 3216.9& 2585.9\\
$m_{\te_L}$                 &  765.6&  845.9& 1036.9& 1532.4\\
$m_{\te_R}$                 & 1460.5& 1402.3& 1688.8& 2343.8\\
$m_{\tnu_{e}}$              &  761.4&  842.1& 1033.8& 1530.3\\
$m_{\ttau_1}$               &  615.9&  680.6&  911.6& 1357.3\\
$m_{\tnu_{\tau}}$           &  622.5&  679.8&  919.7& 1357.3\\
$m_{\tw_1}$                 &  434.3&  640.8&  837.9& 1283.6\\
$m_{\tz_2}$                 & 1610.1&  823.0& 2063.1& 1542.8\\
$m_{\tz_1}$                 &  434.1&  640.4&  837.7& 1283.4\\
$m_h      $                 &  122.5&  119.6&  123.1&  123.4\\
$m_A      $                 &  843.9&  659.3&  628.2&  877.5\\
$m_{H^+}  $                 &  849.9&  666.5&  635.9&  883.9\\
$\mu      $                 & 1734.7&  875.7& 2253.8& 1661.9\\
$a_\mu \times 10^{10}$      &  35.99&  48.21&  18.50&   8.04\\
$B(b\to s\gamma)\times 10^4$&   3.78&   3.65&   3.96&   4.33\\
\hline
\end{tabular}
\caption{
Representative model parameters, and (s)particle masses (in GeVs) for models with
good Yukawa unification.
}
\label{tab:one}
}

\section{Summary and conclusions}
\label{sec:conclusions}

In this work, we analysed Yukawa unification in models with non-universal gaugino mediation of SUSY breaking.
We assumed that all soft SUSY breaking terms vanish at the compactification scale $M_c$ (which is set slightly 
below the usual SUSY GUT scale). The exception being the gaugino and Higgs masses, which were assumed to be 
non-zero and independent at $M_c$. We also considered the cases with vanishing Higgs masses and degenerate 
$M_2$ and $M_3$ values.

We showed that $b - \tau$, and even $t - b - \tau$,  unification can be satisfied
simultaneously with experimental constraints on $b \rightarrow s \gamma $ and  $a_\mu$.
This typically happens for $|M_1 | \gg |M_2 | , | M_3 |$. The large values of $M_1$ are needed in order to 
have a neutralino LSP in a substantial part of the parameter space. This is necessary especially in the large 
$\tan \beta$ region where Yukawa unification can be achieved.

Yukawa unification strongly prefers $\mu M_3 < 0$ just as the mSUGRA scenario or SO(10) motivated models~\cite{bdr, 
baer_so10, Auto:2003ys}. The advantage of non-universal gaugino mediation is the possibility of achieving Yukawa coupling 
unification with a relatively light spectrum. This, on the other hand, provides a sizeable SUSY contribution to the muon
anomalous magnetic moment (shown by Table~\ref{tab:one}), while the SUSY contribution to $a_\mu$ in most of other scenarios is
negligible~\cite{bdr, baer_so10,tobe_wells, Auto:2003ys}.
The relic density of neutralinos from resulting  region of SUSY parameter space is typically below the expectations
from cosmological observations. For recent discussion of neutralino relic density in other frameworks for Yukawa coupling 
unification, see Refs.~\cite{Auto:2003ys, drrr_dm, nonuniv_dm, quasi_dm}.

\section*{Acknowledgments}

We would like to thank H. Baer, S. Raby, K. Tobe, T. Bla\v zek and X. Tata 
for fruitful discussions.
R.D. was supported, in part, by the U.S.\ Department of Energy, Contract 
DE-FG03-91ER-40674 and the Davis Institute for High Energy Physics.
The research of C.B. was supported by the U.S. Department of Energy under 
contract number DE-FG02-97ER41022.


\begin{thebibliography}{99}

\bibitem{su4xsu2xsu2}
J. Pati and A. Salam, \prd{8}{1973}{1240}; \\
J. Pati and A. Salam, \prd{10}{1974}{275}.

\bibitem{su5}
H. Georgi and S. Glashow, \prl{32}{1974}{438}.

\bibitem{so10}
H. Georgi, Particles and Fields, Proceedings of the APS Div.
of Particles and Fields, ed C. Carlson, p. 575 (1975); \\
H. Fritzsch and P. Minkowski, {\it Ann. Phys.} {\bf 93}, (1975) 193.

\bibitem{eft} H. Georgi, H.R. Quinn and S. Weinberg, \prl{33}{1974}{451}; \\
S. Weinberg, \plb{91}{1980}{51}.

\bibitem{susygut}
S. Dimopoulos, S. Raby and F. Wilczek, \prd{24}{1981}{1681}; \\
S. Dimopoulos and H. Georgi,  \npb{193}{1981}{150}; \\
L. Ibanez and G.G. Ross, \plb{105}{1981}{439}; \\
N. Sakai, \zpc{11}{1981}{153}; \\
M.B. Einhorn and D.R.T. Jones, \npb{196}{1982}{475}; \\
W.J. Marciano and G. Senjanovic, \prd{25}{1982}{3092}.

\bibitem{gutexp}
U. Amaldi, W. de Boer and H. F\"urstenau, \plb{260}{1991}{447}; \\
J. Ellis, S. Kelly and D.V. Nanopoulos,  \plb{260}{1991}{131}; \\
P. Langacker and M. Luo, \prd{44}{1991}{817}.

\bibitem{nothreshcorr} For early discussion of fermion masses in grand unified theories, see \\
T. Banks, \npb{303}{1988}{172}; \\
M. Olechowski and S. Pokorski, \plb{214}{1988}{393}; \\
S. Pokorski, \npb{13}{1990}{606} (Proc. Supp.); \\
B. Ananthanarayan, G. Lazarides and Q. Shafi, \prd{44}{1991}{1613};\\
Q. Shafi and B. Ananthanarayan, ICTP Summer School lectures (1991); \\
S. Dimopoulos, L.J. Hall and S. Raby, \prl{68}{1992}{1984}; \\
ibid \prd{45}{1992}{4192}; \\
G. Anderson et al., \prd{47}{1993}{3702}; \\
B. Ananthanarayan, G. Lazarides and Q. Shafi, \plb{300}{1993}{245}; \\
G. Anderson et al.,  \prd{49}{1994}{3660}; \\
B. Ananthanarayan, Q. Shafi and X.M. Wang, \prd{50}{1994}{5980}.

\bibitem{threshcorr}  
L.J. Hall, R. Rattazzi and U. Sarid, \prd{50}{1994}{7048}; \\
R. Hempfling, \prd{49}{1994}{6168};  \\
M. Carena, M. Olechowski, S. Pokorski and C. E. Wagner, \npb{426}{1994}{269}; \\
T. Blazek, S. Pokorski and S. Raby, \prd{52}{1995}{4151}; \\
R. Rattazzi and U. Sarid, \prd{53}{1996}{1553}; \\
D. M. Pierce, J. A. Bagger, K. T. Matchev and R. J. Zhang, \npb{491}{1997}{3}.

\bibitem{muon_g-2} 
H.N. Brown et al., \prl{86}{2001}{2227}; \\
U. Chattopadhyay and P. Nath, \prd{53}{1996}{1648}.

\bibitem{bdr}
T. Bla\v{z}ek, R. Derm\' \i \v sek and S. Raby, \prl{88}{2002}{111804}; \\
ibid \prd{65}{2002}{115004}; \\
S. Raby, talk presented at SUSY 2001, Dubna, Russia, June 2001, \hepph{0110203}; \\
R. Derm\' \i \v sek, talk presented at SUSY 2001, Dubna, Russia, June 2001, \hepph{0108249}.

\bibitem{baer_so10}
H. Baer, M. Diaz, J. Ferrandis and X. Tata, \prd{61}{2000}{111701};\\
H. Baer and J. Ferrandis,  \prl{87}{2001}{211803}.

\bibitem{Auto:2003ys}
D. Auto, H. Baer, C. Bal\'azs, A. Belyaev, J. Ferrandis and X. Tata,
\hepph{0302155}.

\bibitem{tobe_wells}
K. Tobe and J. D. Wells, \hepph{0301015}.

\bibitem{yukawa_nonuniversalities}
U. Chattopadhyay and P. Nath, \prd{65}{2002}{075009}; \\ 
S. Komine and M. Yamaguchi, \prd{65}{2002}{075013}.

\bibitem{yukawa_other}
M. E. Gomez, G. Lazarides and C. Pallis, \npb{638}{2002}{165}; \\
B. Bajc, G. Senjanovic and F. Vissani, \hepph{0210207}; \\
J. Ferrandis, \hepph{0211370}.

\bibitem{proton}
See for example,
T. Goto and T. Nihei, \prd{59}{1999}{115009}; \\
K. S. Babu, J. C. Pati and F. Wilczek, \npb{566}{2000}{33}; \\
R. Derm\' \i \v sek, A. Mafi and S. Raby, \prd{63}{2001}{035001}; \\
G. Altarelli, F. Feruglio and I. Masina, \jhep{040}{2000}{0011}; \\
H. Murayama and A. Pierce, \prd{65}{2002}{055009}; \\
B. Bajc, P. F. Perez and G. Senjanovic, \prd{66}{2002}{075005}; \\
D. Emmanuel-Costa and S. Wiesenfeldt, \hepph{0302272}.

\bibitem{kawamura}
Y. Kawamura, \ptp{105}{2001}{999}; ibid \ptp{105}{2001}{691}.

\bibitem{hall_nomura}
L. Hall and Y. Nomura, \prd{64}{2001}{055003}.

\bibitem{other_models}
G. Altarelli and F. Feruglio, \plb{511}{2001}{257};\\
A.B. Kobakhidze, \plb{514}{2001}{131};\\
A. Hebecker and J. March-Russell, \npb{613}{2001}{3};\\
N. Haba, Y. Shimizu, T. Suzuki and K. Ukai, \ptp{107}{2002}{151};\\
T. Li, \plb{520}{2001}{377}.

\bibitem{so10in5d}
R. Derm\' \i \v sek and A. Mafi, \prd{65}{2002}{055002}.

\bibitem{so10in6d}
L. Hall, Y. Nomura, T. Okui and D. Smith, \prd{65}{2002}{035008}; \\
T. Asaka, W. Buchm\"uller and L. Covi, \plb{523}{2001}{199}.

\bibitem{so10in5d_fsu5}
For an example of a model with flipped SU(5) boundary conditions for gaugino masses, see \\
S.M. Barr and I. Dorsner, \prd{66}{2002}{065013}.

\bibitem{other_susy_br}
L. Hall and Y. Nomura, \prd{66}{2002}{075004}; \\
N. Haba, Y. Shimizu arXiv:hep-ph/0210146.

\bibitem{ginoMSB1}
D. E. Kaplan, G. D. Kribs and M. Schmaltz, \prd{62}{2000}{035010}; \\
Z. Chacko, M. A. Luty, A. E. Nelson and E. Ponton, \jhep{0001}{2000}{003}.

\bibitem{ginoMSB2}
M. Schmaltz and W. Skiba, \prd{62}{2000}{095004}; \\
ibid \prd{62}{2000}{095005}.

\bibitem{gaugino_Baer}
H. Baer, M. Diaz, P. Quintana and X. Tata, \jhep{0004}{2000}{016}; \\
H. Baer, A. Belyaev, T. Krupovnickas and X. Tata, \prd{65}{2002}{075024}.

\bibitem{nonuniversal_gaugino}
H. Baer, C. Bal\' azs, A. Belyaev, R. Derm\' \i \v sek, A. Mafi and A. Mustafayev, \jhep{0205}{2002}{061}.

\bibitem{RGEs}
S. P. Martin and M. T. Vaughn, \prd{50}{1994}{2282}.

\bibitem{isajet} ISAJET, by F. Paige, S. Protopopescu, H. Baer and X. Tata,
hep-ph/0001086 (2000).  

\bibitem{softsusy} B. Allanach, Comput. Phys. Commun. {\bf 143},  
305 (2002).  

\bibitem{spheno} W.~Porod, 
[arXiv:hep-ph/0301101].

\bibitem{suspect} Suspect, by A. Djouadi, J. Kneur and G. Moultaka,  
hep-ph/0211331 (2002). 

\bibitem{kraml} B. Allanach, S. Kraml and W. Porod,  
[arXiv:hep-ph/0302102]. 

\bibitem{Baer:2002gm}
H.~Baer, C.~Bal\'azs, A.~Belyaev, J.~K.~Mizukoshi, X.~Tata and Y.~Wang,
JHEP {\bf 0207}, 050 (2002)
[arXiv:hep-ph/0205325].

\bibitem{Baer:2001kn}
H.~Baer, C.~Bal\'azs, J.~Ferrandis and X.~Tata,
Phys.\ Rev.\ D {\bf 64}, 035004 (2001)
[arXiv:hep-ph/0103280].

\bibitem{lep2} See {\it e.g.} R.~Barate {\it et al.}  [ALEPH Collaboration],
\plb{499}{2001}{67}.

\bibitem{Baer:2002ay}
H.~Baer, C.~Bal\'azs, A.~Belyaev, J.~K.~Mizukoshi, X.~Tata and Y.~Wang,
arXiv:hep-ph/0210441.

\bibitem{Bennett:2003bz}
C.~L.~Bennett {\it et al.},
arXiv:astro-ph/0302207.

\bibitem{drrr_dm}
R. Derm\' \i \v sek, S. Raby, L. Roszkowski and R. Ruiz de Austri, 
in preparation.

\bibitem{nonuniv_dm}
U. Chattopadhyay, A. Corsetti and P. Nath, \prd{66}{2002}{035003}. 

\bibitem{quasi_dm}
C. Pallis and M.E. Gomez, arXiv:hep-ph/0303098.
 
\end{thebibliography}
\end{document}